\documentclass[12pt,b4paper,hyper]{JHEP3}
\usepackage{amsmath}
\usepackage{amscd}
\usepackage{amsfonts}
\usepackage{amssymb}
\usepackage{amsthm}

\usepackage{cite}
\usepackage{epsfig}

\newcommand{\DeA} {^{\displaystyle \bullet} \De}

\newcommand {\ft} {\footnote}
\newcommand{\rf}[1]{(\ref{#1})}
\newcommand{\fig}[1]{fig.~\ref{#1}}
\newcommand{\eq}  [1]{\begin{equation} #1 \end{equation}}

\newcommand{\muu}[1]{\begin{multline} #1 \end{multline}}
\newcommand{\alid}[1]{\begin{aligned} #1 \end{aligned}}

\newcommand {\tr}   {\mathop {\rm Tr}}

\newcommand{\rk}{\mathop{\rm rank}\nolimits}
\newcommand{\vol}{\mathop{\rm Vol}\nolimits}

\renewcommand{\det}{\mathop{\rm det}\nolimits}

\newcommand{\SO}{\mathop{\rm SO}\nolimits}
\newcommand{\Spin}{\mathop{\rm Spin}\nolimits}
\newcommand{\GL}{\mathop{\rm GL}\nolimits}
\newcommand{\SL}{\mathop{\rm SL}\nolimits}
\newcommand{\bigw}{\mathop{\bigwedge}\nolimits}

\newcommand{\Diff}{\mathop{\rm Diff}}

\newcommand{\Un}{\mathop{\rm {}U}\nolimits}

\newcommand{\sO}{\mathfrak{so}}

\newcommand{\sL}{\mathfrak{sl}}

\newcommand{\End}{\mathop{\rm End}}
\newcommand{\Hom}{\mathop{\rm Hom}}

\newcommand{\td}  {\mathop{\rm Td}}
\newcommand{\ch}  {\mathop{\rm Ch}}

\renewcommand{\Im} {\mathop{\rm Im}}
\renewcommand{\Re}    {\mathop{\rm Re}}

\newcommand {\f}  {\frac}
\newcommand {\fr}   {\tfrac}

\renewcommand{\geq}{\geqslant}
\renewcommand{\leq}{\leqslant}

\renewcommand {\d}    {\partial}
\newcommand {\dbar}   {\bar \partial}

\newcommand {\al} {\alpha}

\newcommand {\be} {\beta}

\newcommand {\de} {\delta}
\newcommand {\ve}  {\varepsilon}
\newcommand{\e}{\epsilon}

\renewcommand {\l} {\lambda}

\renewcommand {\th}   {\theta}

\newcommand {\om} {\omega}

\newcommand {\Om} {\Omega}
\newcommand {\Ga} {\Gamma}

\newcommand {\De} {\Delta}

\newcommand {\CalA} {\mathcal A}

\newcommand {\CalF} {\mathcal F}

\newcommand {\CalJ} {\mathcal J}

\newcommand {\N}    {\mathcal N}

\newcommand {\CalL} {\mathcal L}

\newcommand {\CalM} {\mathcal M}
\newcommand {\CalP} {\mathcal P}

\newcommand {\BR}   {\mathbb R}
\newcommand {\BZ}   {\mathbb Z}
\newcommand {\BC}   {\mathbb C}

\newcommand {\Unity}{\mathbf 1}

\newcommand{\g}{\mathfrak{g}}

\newcommand {\ti}  {\tilde}

\newcommand {\la} {\left \langle}
\newcommand {\ra} {\right \rangle}
\newcommand {\lb} {\left (}
\newcommand {\rb} {\right )}

\newcommand{\ket}[1]{\left |  #1 \right \rangle}

\newcommand{\omA} {^{\displaystyle \bullet} \Om}
\newcommand{\omB} {\Om^{\displaystyle \bullet}}
\newcommand{\omC} {_{\displaystyle \bullet}\Om}
\newcommand{\omD} {\Om_{\displaystyle \bullet}}

\newcommand{\omAB} {\overset{\displaystyle \bullet}{\Om}}
\newcommand{\omCD} {\underset{\displaystyle \bullet}{\Om}}
\newcommand{\omAC} {{\bullet}  \Om}
\newcommand{\omBD} {\Om \bullet}

\newcommand{\DeAB} {\overset{\displaystyle \bullet}{\De}}
\newcommand{\DeCD} {\underset{\displaystyle \bullet}{\De}}
\newcommand{\DeAC} {{\bullet}  \De}
\newcommand{\DeBD} {\De \! \bullet}

\title{The Hitchin functionals and the topological B-model at one loop }

\author{Vasily Pestun\thanks{On leave of absense from ITEP, 117259, Moscow, Russia}\\
Princeton University, Princeton NJ, 08544 }

\author{Edward Witten\\
Institute for Advanced Study, Princeton NJ, 08540}

\preprint{ ITEP-TH-09/05 \\PUPT-2154
\\ \hepth{yyyymmdd}}

\abstract{
  The quantization in quadratic order of the Hitchin functional,
which defines by critical points a Calabi-Yau structure on a
six-dimensional manifold, is performed. The conjectured relation
between the topological $B$-model and the Hitchin functional is
studied at one loop. It is found that the genus one free energy of
the topological $B$-model \emph{disagrees} with the one-loop free
energy of the minimal Hitchin functional. However, the topological
$B$-model  \emph{does agree} at one-loop order with  the
\emph{extended Hitchin functional}, which also defines by critical
points a \emph{generalized} Calabi-Yau structure.
The dependence of the one-loop result on a background metric is
studied, and a gravitational anomaly is found for both the
$B$-model and the extended Hitchin model. The anomaly reduces to a
volume-dependent factor if one computes for only Ricci-flat Kahler
metrics.
}

\begin{document}
\section{Introduction}
In~\cite{Dijkgraaf:2004te,Gerasimov:2004yx,Nekrasov:2004vv} it was
suggested that the Hitchin functional $H(\rho)$ defined
in~\cite{Hitchin} for real 3-forms on a 6-dimensional manifold $X$
is related to the topological
$B$-model~\cite{Witten:1991zz,Witten:1988xj,Witten:1988ze,Bershadsky:1993cx}
with the same target space $X$. Moreover
in~\cite{Dijkgraaf:2004te} a qualitative relation was proposed,
which states that the partition function of the Hitchin model on
$X$ is the Wigner transform of the partition functions of $B$ and
$\bar B$ topological strings on $X$. See
also~\cite{Ooguri:2004zv,Vafa:2004qa,Aganagic:2004js}. That
relation was tested in~\cite{Dijkgraaf:2004te} at the classical
level by evaluating the integral Wigner transform by the saddle
point method.

The aim of the current work is to perform quantization of the
Hitchin functional at the quadratic or one-loop order and to test
the relation of it to the topological B-strings. It is
known~\cite{Bershadsky:1993cx} that the one-loop contribution to
the partition function of the $B$-model is given by the product of
the holomorphic Ray-Singer $\dbar$-torsions $I^{RS}_{\dbar,p}$ for
the bundles of holomorphic $p$-forms  $\Om^{p,0}(X)$ on $X$:
    \eq{\label{actors} F_1^B  =  \log \prod_{p=0}^{n} (I^{RS}_{\dbar,p})^{(-1)^{p+1} p }.}
The holomorphic Ray-Singer torsion for the $\dbar$-complex of the
holomorphic vector bundle $\Om^{p,0}$ is defined as the
alternating product of the determinants of Laplacians\ft{Here
$\det' \De_{pq}$ denotes the $\zeta$-function regularized product
of non-zero eigenvalues of the Laplacian $\De_{pq}$.} acting on
$(p,q)$ forms:
    \eq{ I^{RS}_{\dbar,p} (\Om^{p,0}) = \lb \prod_{q=0}^n (\det' \De_{pq})^{(-1)^{q+1} q } \rb^{1/2}.}
In terms of determinants, the formula for $F_1^B$
is~\cite{Bershadsky:1993cx}
    \eq{\label{F_1B} F_1^B  = \f 1 2 \log \prod_{p=0}^{n} \prod_{q=0}^n (\det' \De_{pq})^{(-1)^{p+q} pq}.}
Alternatively, the one-loop factor in the $B$-model partition is
\eq{\label{bactors} Z_{B,1-loop}=\exp(-F_1^B)={\f{I_1} { I_0^3}},}
where we have used the fact that by Hodge duality, $I_p=I_{3-p}$.

In the present work, we show that after appropriate gauge fixing
at the one-loop level, the partition function for the Hitchin
functional $Z_{H,1-loop}$  also reduces to a product of
holomorphic Ray-Singer torsions $I_{\dbar,p}^{RS}$, but with
different exponents from  ~\rf{bactors}.

We suggest resolving this discrepancy by considering instead of
$H(\rho)$ the \emph{extended Hitchin functional} $H_{E} (\phi)$
for polyforms $\phi$, which defines the generalized Calabi-Yau
structure\cite{Hitchin:2,GCS} on $X$. (A polyform is simply a sum of differential
forms of all odd or all even degrees.)  The partition function of
the \emph{extended Hitchin functional} will be called $Z_{HE}$.
The functional $H_{E}(\phi)$ was studied by Hitchin in his
subsequent work~\cite{Hitchin:2,GCS}. For CY manifolds~$X$ with
$b_1(X) = 0$, the critical points and classical values of both
functionals coincide. Hence the \emph{extended Hitchin functional}
$H_{E}(\phi)$ will also fit the check on classical level
in~\cite{Dijkgraaf:2004te} in the same way as $H(\rho)$.
 However the quantum fluctuating degrees of freedom are different for $H_{E}(\phi)$ and $H(\rho)$.

The precise statement about the relation between the full
partition function of topological $B$-model and Hitchin functional
still has to be clarified. In this paper, we will show that the
one-loop contribution to the partition function of \emph{extended
Hitchin theory} is equal to the one-loop contribution to the
partition function of the $B$-model
    \eq{\label{res} \boxed{F_{HE,1-loop} = F_{B,1-loop}}}
The integral Wigner transform in~\cite{Dijkgraaf:2004te}, computed
by the steepest descent method is equivalent at the classical
level to the Legendre transform studied in~\cite{Nekrasov:2004vv}.
The relation shall still hold for the first order correction. Thus
the conjecture in \cite{Dijkgraaf:2004te} that $Z_{H} =
Wigner[Z_{B} \otimes \bar Z_{B}]$ appears to contradict our
result, since the extended Hitchin functional agrees with
$Z_{B,1-loop}$ rather than its absolute value squared.

However, this apparent discrepancy has been explained to us by A.
Neitzke and C. Vafa, as follows.  Although formally it appears
that the $B$-model partition function computed in genus $g$ should
be holomorphic, this is in fact not so as there is a holomorphic
anomaly \cite{Bershadsky:1993cx}.  In genus one, the role of the
holomorphic anomaly is particularly simple; the genus one
computation is symmetrical between the $B$-model and its complex
conjugate, and apart from contributions of zero modes, one has
$F^B_1=F^{B,hol}_1+c.c.$, where $F^{B,hol}_1$ is a holomorphic
function of complex moduli.  Hence
$\exp(-F^B_1)=|\exp(-F^{B,hol}_1)|^2$.  So our result \rf{res}
means that $Z_{HE,1-loop}=\exp(-F_{HE,1-loop})$ is equal to
$|\exp(-F_1^{B,hol})|^2$.   Further, $\exp(-F_1^{B,hol})$ must be
understood as the one-loop factor in $Z_B$ in the conjecture of
\cite{Dijkgraaf:2004te}. With this interpretation, our computation
for the extended Hitchin functional agrees with the conjecture at
one-loop order.

The appearance of the \emph{extended Hitchin functional} for the
generalized CY structures is appealing from the viewpoint of $\N =
1$ string compactifications on 6-folds with
fluxes~\cite{Grana:2004bg,Lindstrom:2004iw,Gurrieri:2002wz,Fidanza:2003zi}.
 A study of the  extended Hitchin functional for generalized complex
 structures, as opposed to ordinary ones, is also natural because the total moduli space of topological
B-strings includes the space of complex structures, but has other
directions as well, corresponding to the sheaf cohomology groups
$H_{\dbar}^q(\bigw^p(TX^{1,0}))$, where the case of $p=q=1$ yields
the Beltrami differentials -- deformations of the complex
structure. See
also~\cite{Kapustin:2003sg,Kapustin:2004gv,Kapustin:2005uy,Zabzine:2004dp,Gurrieri:2002wz,
Zabzine:2005qf,Zucchini:2005rh,Jeschek:2004wy,Chiantese:2004pe,Chiantese:2004pf,Ikeda:2004cm,Jeschek:2004je}.

The paper has the following structure. In section~2, the Hitchin
functionals are quantized at the quadratic order. Section~3
studies the dependence of the result on a background metric, and
shows the presence of  a gravitational anomaly.
In general, the extended Hitchin and $B$-models are not invariant
under a change of Kahler metric.  The dependence on the Kahler
metric does disappear, however, if we only use Ricci-flat Kahler
metrics.  Section~4 concludes the paper.


\section{Quantization of the Hitchin functionals at the quadratic order}

In \cite{Hitchin,Hitchin:2} Hitchin defines two functionals
$H(\rho)$ and $H_E(\phi)$ that will be studied here.

The first functional $H(\rho)$ is defined for real 3-forms $\rho$
on a six-dimensional manifold $X$
    \eq{\label{def_H} H(\rho) = \int_{X} \phi(\rho).}
Here $\phi(\rho)$ is a measure on $X$, defined as a nonlinear
functional of a 3-form $\rho=\frac 1 {3!}\rho_{i_1i_2i_3}dx^{i_1}
\wedge dx^{i_2}\wedge dx^{i_3}$ by
    \eq{ \phi(\rho) = \sqrt{-\l(\rho)} d^6 x,}
where
    \eq{ \l(\rho) = \f 1 6 \tr K(\rho)^2} and the matrix $K^{b}_{a}(\rho)$
is given by
    \eq{  K^{b}_{a}(\rho) = \f 1 {2!3!}  \e^{b i_2 i_3 i_4 i_5 i_6} \rho_{a i_2
    i_3}\rho_{i_4 i_5 i_6} .}
Thus, $\phi(\rho)$ is the square root of a fourth order polynomial
constructed from the components of the 3-form $\rho$. As Hitchin
showed, each 3-form with $\l(\rho)< 0$ defines an almost complex
structure on $X$. This almost complex structure is integrable if
$d\rho=0$ and $d \hat \rho =0$, where $\hat\rho$ is a certain
3-form constructed as a nonlinear function of $\rho$. In the space
of closed 3-forms $\rho$ in a given cohomology class, the
Euler-Lagrange equations are $d\hat \rho=0$. A critical point or
classical solution hence defines an integrable complex structure
$I$ on $X$ together with a holomorphic (3,0)-form $\Om = \rho+ i
\hat\rho$.  We will refer to an integrable complex structure with
such a holomorphic $(3,0)$-form as a Calabi-Yau or CY structure.

To implement the restriction of $\rho$ to a particular cohomology
class $[\rho_0]$, we write $\rho = \rho_0+db$, where $\rho_0$ is
any three-form in the chosen cohomology class, and $b$ is a
two-form that will be quantized.  Of course, $b$ is subject to a
gauge-invariance $b\to b+d\lambda$.
 It turns out that
the second variation of the Hitchin functional $H(\rho)$ on the
manifold $X$ is nondegenerate as a function of $b$ modulo the
action of the group of diffeomorphisms $\Diff_0(X)$ plus gauge
transformations. For $[\rho_0]$ in  a certain open set in
$H^3(X,\BR)$, the Hitchin functional has a unique extremum modulo
gauge transformations and therefore produces a local mapping from
$H^3(X,\BR)$ into the moduli space of CY structures on $X$.

The \emph{extended Hitchin functional}~\cite{Hitchin:2} is
designed to produce \emph{generalized complex}
structures~\cite{GCS} on $X$. To construct the extended
functional, Hitchin replaces the 3-form $\rho$ by a  polyform
$\phi=\rho_1 + \rho_3 + \rho_5$, which is a sum of all odd forms.
Any such polyform, if nondegenerate, defines a generalized almost
complex structure on $X$ given by a ``pure
spinor''\cite{MR0060497,Berkovits:2004tw,Grange:2004ah} $\phi + i
\hat \phi$ with respect to $\Spin(TX,TX^*)$.
 If in addition $d \phi =0$ and $d \hat \phi=0$, then
this generalized almost complex structure is integrable. If
$b_1(X)=0$, then all generalized complex structures coincide with
the usual ones. This will be the case under consideration in the
present work. In this case, the critical points of the extended
Hitchin model produce usual Calabi-Yau structure on $X$, but the
fluctuating degrees of freedom are  different, and therefore these
two theories are different on a quantum level.

For  the original Hitchin model, we write $\rho =\rho_0+db$, and
obtain the partition function as a path integral over $b$,
    \eq{  Z_{H}([\rho_0]) = \int Db \, e^{-H(\rho)}}
where $b$ is a two form.

To study this functional at the one-loop level, one first needs to
expand $H(\rho)$ around some critical point $\rho_c$, which
describes a complex structure $I$ on  $X$. This complex structure
will be used to define the second variation of the functional.

Using the algebraic properties of $\phi(\rho)$, it is easy to
compute the variation \cite{Hitchin}. The first variation is given
by
    \eq{ \de \phi(\rho)  = \de \rho  \wedge \hat \rho.}
The second variation can be written in terms of the complex
structure~$J$ introduced by Hitchin~\cite{Hitchin} on the space of
3-forms $W = \bigw^3(\BR^6)\otimes \BC$. This complex
structure~$J$ commutes with the Hodge decomposition of $W$ with
respect to $I$, and has the following eigenvalues
\eq{\label{J_values}  \alid{ &J =+ i \quad \text{on} \quad W^{3,0}, W^{2,1}\\
                             &J =- i \quad \text{on} \quad W^{1,2}, W^{0,3}.}}
The second variation of the Hitchin functional $H(\rho)$ is
    \eq {\label{secondH}  H^{(2)}(\rho_c,\de \rho) = \de^2_{\rho} H(\rho) = \int_X \de \rho \wedge J \de
    \rho.}
Hence the one-loop quantum partition function for the Hitchin
functional~\rf{def_H} is given by
    \eq{  Z_{H,1-loop}([\rho_c]) = \int Db \, e^{ -H^{(2)}(\rho_c,db)  },}
where $\rho_c$ is a critical point of $H(\rho)$, which defines a
complex structure on $X$.

In terms of the Hodge decomposition of the two-form $b$ with
respect to the complex structure $I(\rho_c)$ on $X$,
    \eq{ b=b_{20}+b_{11}+b_{02},}
one gets\ft{The quadratic term of the action~\rf{def_S} has already appeared in the
Kodaira-Spencer theory of
gravity~\cite{Bershadsky:1993cx,Dijkgraaf:2004te,Gerasimov:2004yx}
after an appropriate gauge fixing.}
    \eq{\label{def_S} H^{(2)}(\rho_c,db) = \int_X db \wedge J db =
    \int_X \d b_{11} \wedge \dbar b_{11}.}
The terms with $d b_{20}$ and $d b_{02}$ vanish. The  form
$db_{20}$ lies entirely in the $i$-eigenspace of $J$ and therefore
$J db_{20}= i db_{20}$, but the wedge product of two exact forms
gives zero after integration over $X$.

The reason that $b_{20}$ and $b_{02}$ do not appear in the action
is that they can be removed by a diffeomorphism.  Under a
diffeomorphism generated by a vector field $V$, any 3-form $\rho$
transforms as $\delta\rho =(i_Vd +di_V)\rho$ (where $i_V$ is the
action of contraction with $\rho$.)  For $d\rho=0$, the
transformation is just $\delta\rho = d(i_V\rho)$.  As we have
taken $\rho=\rho_0+db$, we can take the    transformation of $b$
under an infinitesimal diffeomorphism to be simply $\delta
b=i_V\rho$.  For any $b$, there is a unique choice of $V$ that
will set to zero $b_{20}$ and $b_{02}$; no determinant arises in
this process, as there is no derivative in the transformation law
$\delta b=i_V\rho$. The path integral will therefore be performed
over $b_{11}$, with the remaining gauge invariance $\delta b_{11}
=\bar\partial \lambda_{10}+\partial \lambda_{01}$.

To understand the quantization of the degenerate term $\int_X \d b
\wedge \dbar b$,  it easier first to turn to the canonical example
of the  theory  of an $n$-form field $f_n$ with action  \eq{ S_0 =
-i\label{def_CS} \int_X f_n \wedge df_n, } studied
in~\cite{MR535228,MR0676337}.\ft{If $n$ odd, then one takes $f_n$
to be bosonic fields, and if $n$ is even, then $f_n$ are
fermionic. Indeed, for bosonic fields $ d(f\wedge f) = df \wedge f
+(-1)^n f \wedge df  = (1+(-1)^n) f \wedge df$. So, if $n$ is
even, than the integrand is the total derivative and the classical
action $S_0=0$.}
 This theory leads to
Ray-Singer torsion~\cite{MR0295381,MR0339293,MR0383463}
 for the de Rham complex, which appears as a resolution of the $d$ operator. See
also~\cite{Witten:1992fb,Witten:1989ip,Axelrod:1989xt,Axelrod:1991vq,Axelrod:1993wr}
and~\cite{MR1482939,MR0676337,MR1852055,Schwarz:2000ct,Schwarz:1992nx}.
Actually, we will only calculate the absolute value of the
partition function of this theory.  The partition function has a
nontrivial phase, the $\eta$-invariant (see~\cite{Witten:1988hf}), which arises because the
action is indefinite. We will not discuss it here because it has
no analog for the Hitchin theories that we will analyze later;
hence, we will simply take the absolute values of all
determinants.

It is instructive to first take the integral formally.  Then, we
turn to the BV
formalism~\cite{Batalin:1981jr,Batalin:1984jr,Henneaux:1989jq,Witten:1990wb,Schwarz:1992nx,Schwarz:2000ct},
which is a powerful formalism for quantizing gauge theories such
as this one with redundant or reducible gauge transformations. The
derivation also shows that, if the cohomology groups $H^k(X)$ are
nontrivial, the product of determinants that we get is best
understood not as a number but as a measure on the space of zero
modes. (Ideally, we would then get a number by integrating over
the space of zero modes, but the meaning of this integral is a
little unclear in the case of ghost zero modes.)  A similar fact
holds in the Hitchin theories, but we will not be so precise in
that case.

 The idea is just
to directly integrate over the fields in the initial functional
integral, and manipulate formally with the volume of zero modes.
So, let us say that $n$ is odd and  $f_n$ is bosonic,  and
consider the functional integral
    \eq{  Z_{fdf} =  \f{1}{ \vol (G) }\int D f_n \,\,\exp( i\int f_n d f_n) .}
    Here ${\rm Vol}(G)$ is the volume of the gauge group.
The operator $d$ acting on $f_n$ has an infinite-dimensional
kernel,  the space of closed forms $\Om^{n}_{closed}$. Formally
evaluating the integral, one gets
    \eq{ Z_{fdf} = {\f 1 {\vol(G)} }\lb \det' d_{n\to n+1}  \rb^{-\f 1 2} \vol
    (\Om^{n}_{closed}),}
where $\det'd_{n}$ is the determinant of the operator $d_n$
mapping $\Om^n/\Om^n_{closed}$ to its image in $\Omega^{n+1}$. The
absolute value of this determinant is conveniently defined as
$|\det'd_n| = \sqrt{ \det'{d^*_{n+1} d_{n}}}$, where
$\Delta'_n=d^*_{n+1}d_n$ maps $\Omega^n/\Om^n_{closed}$ to itself.
The remaining factor is the volume of the kernel, which can be
decomposed as
    \eq{ \vol(\Om^n_{closed}) = \vol(\Om^{n}_{exact}) \vol
    (H^{n}).}
Now, using the map
    \eq{ \Om^{n-1} \overset{d}{\longrightarrow} \Om^{n}_{exact} }
and recalling that the kernel of this map is $\Om^{n-1}_{closed}$, one gets
    \eq{ \vol (\Om^{n}_{exact}) = \vol (\Om^{n-1}/\Om^{n-1}_{closed}) \det' d_{n-1} =
    \f {\vol(\Om^{n-1}) \det' d_{n-1} } { \vol(H^{n-1}) \vol (\Om^{n-1}_{exact}) ,}  }
    where again we factored
    $\vol(\Om^{n-1}_{closed})=\vol(H^{n-1})\vol(\Om^{n-1}_{exact})$.
Going recursively and expressing $\vol(\Om^{n-1}_{exact})$ in
terms of volumes of forms of lower rank one gets finally
    \eq {\label{strpath} Z_{fdf} = {1\over {\rm Vol}(G)}\lb \det' d_{n\to n+1}  \rb^{-\f 1 2} \f { \det' d_{n-1}
    \det'd_{n-3} \cdots }
    {\det'd_{n-2} \det'd_{n-4} \cdots } \, \cdot \, \f { \vol H^n \vol H^{n-2}
    \cdots} {\vol H^{n-1} \vol H^{n-3} \cdots}  \, \cdot \, \f { \vol \Om^{n-1} \vol \Om^{n-3} \cdots}
    {\vol \Om^{n-2}\vol \Om^{n-4} \cdots}
    }
Formally, to make sense of the theory, we must take the volume of
the gauge group to be
    \eq{\label{volg}{\rm Vol{G}}= \f { \vol \Om^{n-1} \vol \Om^{n-3}
     \cdots} {\vol \Om^{n-2}\vol \Om^{n-4}\cdots}.}  (One can justify this intuitively by thinking of the
gauge group as consisting of gauge transformations for fields,
ghosts, ghosts for ghosts, and so on.) Any other definition of the
volume that would lead to a sensible theory would differ from this
one by the exponential of a local integral, which we could anyway
absorb in the definition of the determinants.  (In odd dimensions,
there are no such local integrals in any case.)

Using some elementary facts about determinants that we explain
momentarily, one obtains
    \eq { {\alid Z_{fdf} = I_{RS}^{1/2} \f { \vol H^n \vol H^{n-2}
    \cdots} {\vol H^{n-1} \vol H^{n-3} \cdots} \\
    } .}
Here
    \eq {\label{I_RSCS} I_{RS} = \prod_{p=0,\dots, 2n}
    (\det' d_{p} )^{(-1)^p}  = \lb \prod_{p=0,\dots, 2n+1}
    (\det' \, \Delta_{p})^{(-1)^{p+1}p} \rb^{\f 1 2}}
    is the Ray-Singer torsion.

In general, to define a determinant of a linear operator that maps
between two different vector spaces $H_1$ and $H_2$, one has to
introduce a metric on them. Then, given a linear map $w: H_1 \to
H_2$, the absolute value of the determinant of  $w$ is defined
as\ft{ The metric on $H_1$ and $H_2$ is used to define $w^{*}$ as
$(wx,y) = (x,w^*y)$ for $x \in H_1, \enskip y \in H_2$. After that
the operator $w^*w$ maps inside the same space $w^*w: H_1 \to H_1$
and the determinant is defined as a product of its eigenvalues.
This definition of $\det w$ coincides with the definition by
Gaussian integral of $\int dx\, dy e^{(y,wx)}$, which also
requires a metric $(~,~)$ to induce a measure. } \eq{ |\det w|
\equiv \sqrt{ \det (w^*w). } }

In our problem, the Laplacian operator~$\De_p$ acting on the space
of $p$-forms $\Om^p(X)$ is defined as
    \eq{ \De_p = d d^* + d^* d,}
where $d^*$ is the Hodge conjugated differential $*d*$. This
operator can be conveniently decomposed as follows:
    \eq{ \alid{ \De_p &= \, '\De_p + \De_p', \\
                 '\De_p &= d_{p-1} d^*_p, \quad \De'_p = d^*_{p+1} d_p}}
The Hilbert space of $p$-forms can be Hodge decomposed as
    \eq{ \Om^p = H^p \, \oplus \, '\Om^p \, \oplus \, \Om^{\prime p},}
where $H^p$ consists of harmonic forms, annihilated by both $d$
and $d^*$, $'\Omega^p$ consists of forms annihilated by $d$ but
orthogonal to harmonic forms, and $\Omega'{}^p$ consists of forms
annihilated by $d^*$ but orthogonal to harmonic forms. The nonzero
eigenvalues of $'\De_p$ span $'\Omega^p$, and the nonzero
eigenvalues of $\De_p'$ span $\Omega'{}^p$.  $\Delta_p$ is zero on
$H^p$, equals $'\De_p$ on $'\Omega^p$, and equals $\De_p'$ on
$\Omega'{}^p$.  Consequently, \eq{\det'\Delta_p = \det
\,'\Delta_p\,\det\Delta'_p.} where it is understood that
$\det'\,\Delta_p$ is the product of the nonzero eigenvalues of
$\Delta_p$ and that determinants of $'\Delta_p$ and $\Delta'_p$
are taken in $'\Omega^p$ and $\Omega'{}^p$. We also had
    \eq{ |\det' d_p| =  \lb \det \De'_p \rb^{\f 1 2},} and
     by Hodge symmetry $ \det' \, \De_p = \det' \, \De_{2n+1-p}$.
Finally, $\det\,'\Delta_p=\det\Delta'_{p-1}$.  With these facts,
we have $\prod_{p=0}^{2n+1}\det'\Delta_p^{(-1)^{p+1}p}=
\prod_{p=0}^{2n}\det{}(\Delta'_p)^{(-1)^p}$. Using these relations
one gets~\rf{I_RSCS}.

Now we will repeat this discussion in the BV formalism,
suppressing the role of the de Rham cohomology groups to keep
things simple (for example, they may vanish if $f$ obeys twisted
boundary conditions). In this approach, the gauge symmetries of
the first stage for $f_n$ are the deformations by exact forms $d
f_{n-1}$
    \eq{ \label{gtr1} Q f_n = d f_{n-1}}
So the forms $f_{n-1}$ are the ghosts fields whose ghost number is
$\#gh=1$, and of the opposite statistics to $f_n$.  This gauge
transformation is reducible, since the transformation~\rf{gtr1} is
trivial if $f_{n-1}=df_{n-2}$ for some $f_{n-2}$.  Hence we
introduce ghost number two fields $f_{n-2}$, sometimes called
``ghosts for ghosts'':
    \eq{ Q f_{n-1} = d f_{n-2}}
The process is continued recursively, with
    \eq{\label{BRST} Q f_{k} = d f_{k-1}, \quad 0<k \leq n.}
So the operator $Q$ of the BV complex is mapped to the
differential $d$ acting on forms whose degree ranges from 0 to
$n-1$.

So far we  have introduced differential forms $f_k$, $0\leq k\leq
n$, where $f_n$ are the original fields and the others are ghosts
and ghosts for ghosts.   The statistics of $f_k$ are $(-1)^{k+1}$.
In the BV formalism, we need to introduce antifields, which are
conjugate variables with opposite statistics.  The conjugate
variable to $f_k$ is simply a differential form $f_{2n+1-k}$, of
statistics $(-1)^k=(-1)^{(2n+1-k)+1}$; the antibracket  (odd
Poisson bracket) is defined by the odd symplectic structure
   \eq{ \om(f,g) = \int_X f \wedge g .}
   For all $0\leq k\leq 2n+1$, the ghost number of $f_k$ is $n-k$.

Having now the complete set of fields and antifields, and BRST
transformations \rf{BRST}, we must next find the master action $S$
that coincides with the classical action for the original fields
and  such that $\{S,S\}=0$ and  $\{S,f_k\}=Qf_k$ for all $k$. The
reason that the BV formalism is so convenient for this particular
problem is that the master action is very simple: the BV master
action $S$ is
    \eq{\label{def_Master} S = -i\sum_{k=0}^{n} \int f_k \wedge d f_{2n-k}.}
It has the same form as the classical action $S_0$, except that
the fact that the constraint has degree $n$ is dropped.  This fact
has been used in analyzing Chern-Simons perturbation
theory~\cite{Axelrod:1991vq} and has a powerful analog for string
field theory~\cite{Bochicchio:1986zj}.

The resulting BV complex can be drawn as in \fig{realbv}.
\EPSFIGURE{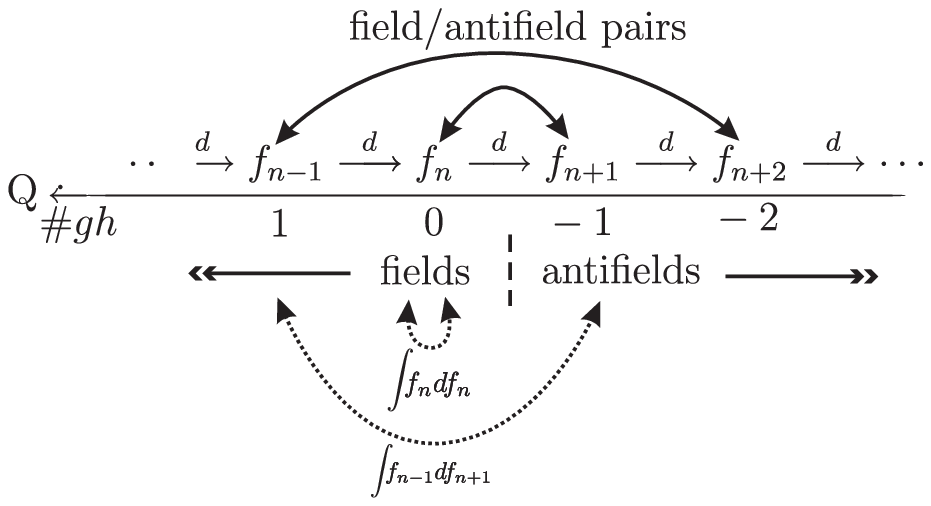}{\label{realbv} The BV complex for $\int_X
df_n \wedge f_n$. Ghost number increases from right to left; the
action of $Q$ increases the ghost number by 1.}

The last step of the BV formalism is to choose \emph{a gauge
fixing fermion $\Psi$}, or, alternatively, to choose a Lagrangian
submanifold in the function space of fields~$f_k$.
A convenient choice of   Lagrangian submanifold $L$ is  the space
of $d^{*}$ closed forms\ft{For $d^*$ closed forms $f,g$ one has,
as we are neglecting the de Rham cohomology, $f=d^*h$ for some
$h$.  Hence $\int_X d^*h \wedge g =0$ by integration by parts,
showing that the condition that the fields be annihilated by $d^*$
gives a Lagrangian submanifold of the space of fields.}
    \eq{L: \quad d^* f_k =0.}
By the Hodge decomposition (and our assumption of ignoring the de
Rham cohomology), this condition is equivalent to restricting the
fields $f_k$ to the orthocomplement of the space of gauge
variations $Qf_k=d f_{k-1}$.  In all subsequent examples, we will
similarly pick our Lagrangian submanifold to be the
orthocomplement of the space of gauge variations.

After the restriction to the Lagrangian submanifold, the partition
function of the theory with quadratic action~\rf{def_Master} is
equal to the alternating ratio of determinants of the differential
operators $d_k$ mapping in the complex~\rf{def_Qcomplex}
    \eq{\label{def_Qcomplex} f_0 \overset{d} {\longrightarrow} f_1 \overset{d}{\longrightarrow}
    f_2 \overset{d}{\longrightarrow} \dots \overset{d}{\longrightarrow} f_n, }
in the function space of differential forms
$\Om^p(X)/\Omega^p(X)_{closed}.$ Indeed, the gauge fixing
condition $d^*_k f_k =0$ eliminates the kernels of the operators
$d$ in~\rf{def_Master}, given that we assume the cohomology groups
to be trivial. By a derivation that we have already seen, the
alternating product of determinants of $d_n$ gives the following
product:
    \eq {Z_{fdf} = \lb (\det \De_n')^{-1/4} \prod_{k=1}^{n} (\det
            \De_{n+k}')^{(-1)^k} \rb^{(-1)^{n+1}} = \lb \prod_{k=0}^{k=d-1} (\det{ \De'_k})^{
            (-1)^k}\rb^{\f 1 4 } = I_{RS}^{ \f 1 2 }}
For the theory $S = \int_X f \wedge dg $, with a doubled set of
variables, one has $Z_{f dg} =I_{RS}$.


\bigskip

\underline{The quadratic term of the Hitchin functional}

Now consider the quadratic term of the Hitchin functional~\rf{def_S}
    \eq {S = \int_X \d b_{11} \wedge \dbar b_{11}.}
To define the determinants of Laplacians that will appear in the
evaluation of the partition function, one has to  introduce a
metric on  $X$. The dependence on it will be analyzed later. We
will use some  Kahler metric $g$ compatible with the complex
structure.

The Hodge diamond on ~\fig{hodge},
\EPSFIGURE{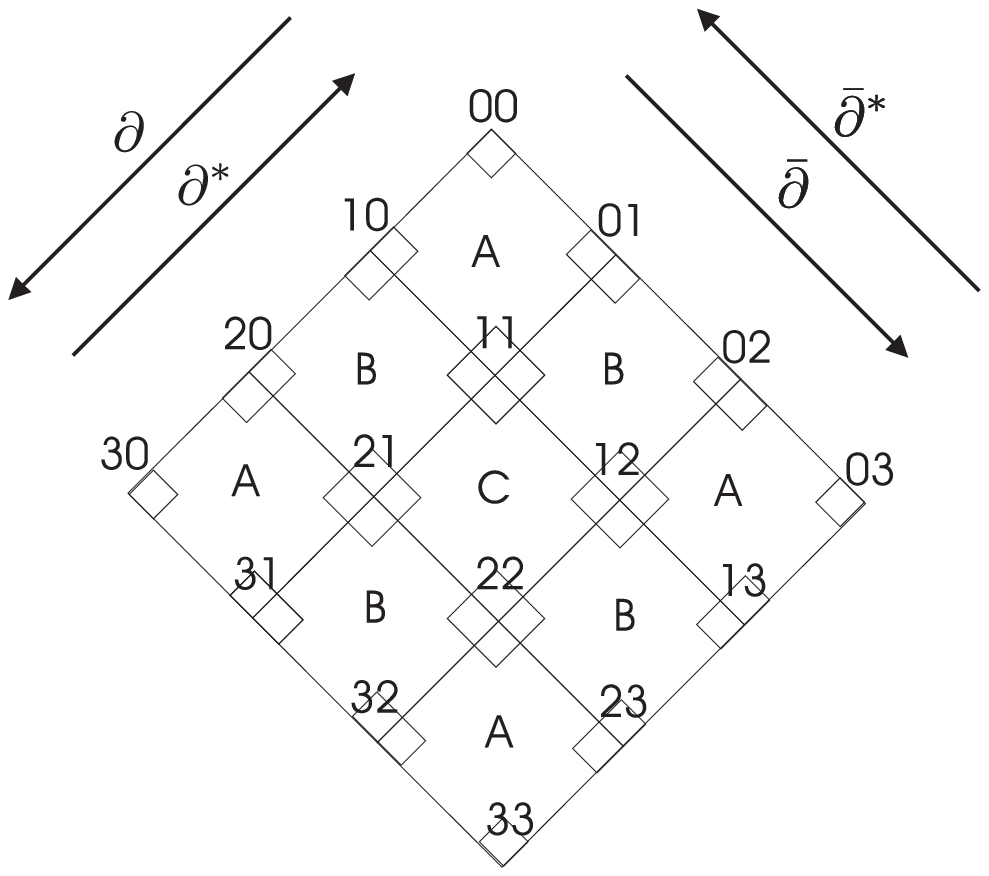}{\label{hodge} The Hodge diamond for a CY
3-fold. The type of forms is shown by label at vertices. The
differential operators $\d, \dbar, \d^*, \dbar^*$ act along
diagonal edges. The labels $A,B,C$ are associated with 9 square
faces and are used to denote corresponding determinants of
Laplacians.  At each vertex, we have drawn some small squares,
which will be used later to depict the Hodge decomposition. The
meaning of the little squares is obtained later.} which is made of
forms of different type $(p,q)$, can be sliced either into $\d$ or
$\dbar$ -complexes along the diagonal lines.

On a Kahler manifold, one has the relation~\cite{MR507725}
    \eq{    \De_{\d} = \De_{\dbar} = \f 1 2 \De_d }
where
    \eq{ \alid{ &\De_{\dbar} = \dbar\dbar^* + \dbar^* \dbar,
    \quad  &\De_{\d}      = \d \d^* + \d^* \d, \quad  &\De_{d}       = d d^* + d^*d }}
For a compact Kahler manifold, the cohomology groups  $H^{p,q}$
are the same both for $\dbar$ and $\d$ operator. They can be
represented by the  space of harmonic $(p,q)$ forms.\ft{The form
of type $(p,q)$ is called $\dbar$-harmonic, if it is annihilated
by the Laplacian $\De_{\dbar}$, and the same is for $\De_{\d}$ and
$\De_{d}$.} Harmonic forms provide a unique representative in each
$\dbar$-cohomology class. The same statement holds for $d$ and
$\d$ operators. Since on Kahler manifolds the $\d$ and $\dbar$
Laplacians are equal
    \eq{ \De_{\d} = \De_{\dbar},}
the  $\dbar$ and $\d$ cohomology groups are represented by the
same space of harmonic forms.

\begin{figure} [h]
\centering
\includegraphics{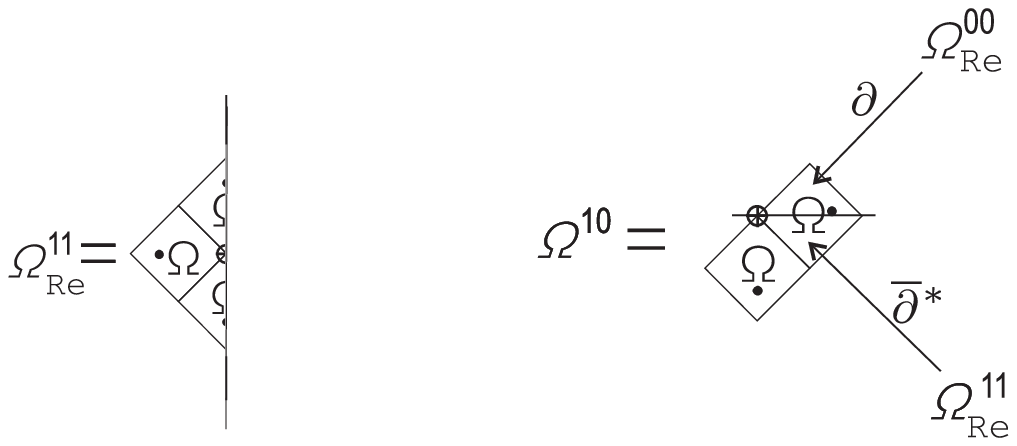}
\caption{To depict the Hodge decomposition of a  real form of type
$(p,p)$, we introduce a vertical line and draw only one half of a
square, symbolizing that the form is self-conjugate.  The Hodge
decomposition of a real form of type $(1,1)$ is sketched on the
left; since ${\overset{ \bullet}{\Om}}{}^{11}$ and
${\underset{\displaystyle \bullet}{\Om}}^{11}$ are real, they are
depicted by little triangles rather than little squares, and as
${\Om \bullet}{}^{11}$ is the complex conjugate of ${{\bullet}
\Om}^{11}$, only one of them is shown. More generally, we will
usually only depict one of each complex conjugate pair of fields.
Another subtlety arises in the Hodge decomposition when
$\Omega^{pq}$ is complex but one of the spaces arising in its
Hodge decomposition is real.  On the right, we show this for
$\Omega^{10}$.  The little square depicting ${\Om \bullet}^{10}$
is divided into upper and lower triangles representing $\partial$
of a real or imaginary $(0,0)$-form, respectively. Alternatively,
they represent images of real forms of type $(0,0)$ and $(1,1)$,
as shown.} \label{real}
\end{figure}

 \EPSFIGURE{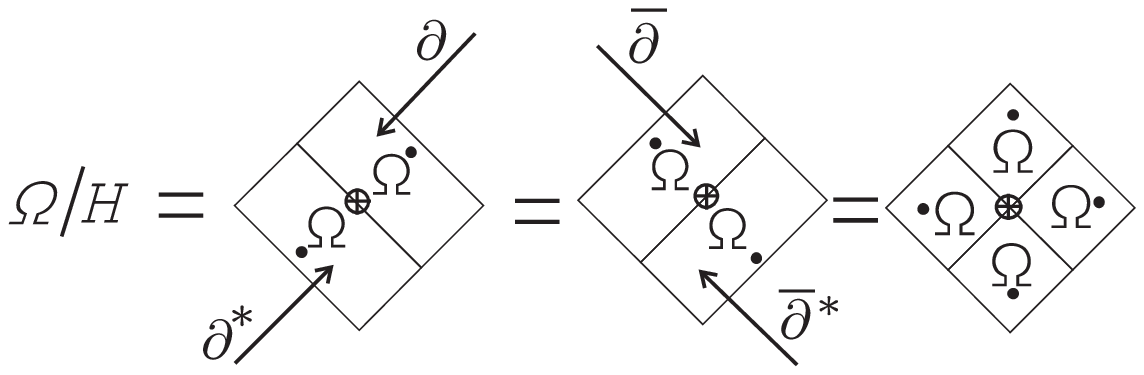}{ \label{squares} Three ways  of
decomposing $\Om^{p,q}/H^{p,q}$. The idea of this figure and \fig{real} is to illustrate the decomposition of Hilbert spaces of forms
into direct sums of subspaces. A Hilbert space, in this case
$\Om^{p,q}/H^{p,q}$, is represented as a square which is a union
of rectangles or smaller squares that correspond to subspaces.  In
this example, for instance, the first decomposition is as the sum
of images of $\partial$ and $\partial^*$. The others are as
indicated. If $p$ or $q$ is 0 or 3, some pieces in these
decompositions are missing, because forms of type $p,q$ cannot be
reached by the action of certain  operators.}

The space $\Om^{p,q}$ of forms of type $(p,q)$ can be Hodge
decomposed either with respect to the operator $\dbar$ or  the
operator $\d$. In the first case, one has
    \eq { \Om^{p,q} =
    H^{p,q}_{\dbar} \, \oplus \, \Im( \dbar( \Om^{p,q-1 })) \, \oplus \, \Im(\dbar^*(\Om^{p,q+1} )),}
and in the second
    \eq { \Om^{p,q} =
    H^{p,q}_{\d} \, \oplus \, \Im( \d( \Om^{p-1,q })) \, \oplus \, \Im(\d^*(\Om^{p+1,q} )).}
To shorten formulas, it is convenient to write
    \eq { \Om = H_{\dbar} \, \oplus \, \omA \, \oplus \, \omD }
and
    \eq { \Om = H_{\d} \, \oplus \, \omB \, \oplus \, \omC.}
The $\bullet$-sign denotes the projection operator on the
corresponding subspace in $\Om^{p,q}$ in accordance with direction
from which this subspace can be obtained by mapping by the $\dbar$
or $\d$ operator, etc., as shown in \fig{squares}. In these
formulas the indices ($p,q$) were suppressed to keep the notation
visually simpler, but they should be kept in mind.

Since $\De_{\d}= \De_{\dbar}$, one can further decompose
$\Om^{p,q} \equiv \Om$. Namely, we define the projection to the
``upper'' and ``lower'' squares
    \eq{\label{updownbullets} \alid{
       \omAB \, := \, \omA \, \cap \, \omB \\
       \omCD \, :=\, \omC \, \cap \,\omD }}
and the ``left'' and ``right'' squares
     \eq{ \alid{
       \omAC \, :=\, \omA \, \cap \,\omC \\
       \omBD \, :=\, \omB \, \cap \,\omD }}
in~\fig{squares}.

That is equivalent to
    \eq{ \alid{
       \omA \, &= &\omAB \, &\oplus \, \omAC \\
       \omB \, &=  &\omBD \, &\oplus \, \omAB \\
       \omC \, &=  &\omCD \, &\oplus \, \omAC \\
       \omD \, &=  &\omCD \, &\oplus \, \omBD  }}
and altogether
    \eq{\label{omdecomp} \alid{
    \Om =H \, \oplus \,         \omAB \, \, \oplus \, \omBD \, {} \oplus \, \, \omCD \,
    \, \oplus \, \, \omAC .}}
    For example, $\Omega$ with a dot above represents the subspace
    of $\Omega$ which is the image of $\partial\bar\partial$
    acting from above, $\Omega$ with a dot to the right represents
    the image of $\partial\bar\partial^*$ acting from the right,
    etc.
Pictorially, we represent this in~\fig{squares} by decomposing
$\Omega$, or rather the orthocomplement to $H$ in $\Omega$, in
terms of four little squares depicted on the right of the figure.
(One could think of $H$ as living at the vertex at the center of
the square.) In general, when we draw a Hodge diamond, each vertex
is associated with such little squares, as shown in~\fig{hodge}.

The decomposition of $\Omega^{p,q}$ simplifies a little if $p$ or
$q$ is equal to $0$ or 3, because then some of the summands in
this decomposition vanish.  This is clear in~\fig{hodge}, where a
vertex on an edge or corner of the Hodge diamond has fewer little
squares attached to it.   Later we will also deal with reality
conditions for forms. Without developing special notation, let us
just summarize the idea in \fig{real}.

Each projection operator from $\Omega$ to one of its ``dotted''
subspaces  commutes with the Laplacian $\De_{\d} = \De_{\bar \d}$.
And of course the Laplacian annihilates $H^{p,q}$. Therefore, the
Laplacian on $\Om^{p,q}$ can be decomposed into the sum defined by
the projection operators $\bullet$
    \eq{ \De  ={\hat 0 }|_{H^{p,q}} \oplus \DeAB \,  \oplus \, \DeAC \,\oplus \, \DeBD \, \oplus \,
    \DeCD.}
    Then the determinant of Laplacian restricted on
$\Om^{p,q}/H^{p,q}$ is  a product
    \eq{ \det' \De = \det (\DeAB) \det (\DeBD )\det
    (\DeCD)\det (\DeAC ).}
    This formula holds for each $(p,q)$. Of course, the Laplacians with $\bullet$-sign do not contain
zero eigenvalues, as all zero modes are contained in $H^{p,q}$.

To recapitulate all this, the Laplacians $\De_{p,q}$ are
associated with the vertices in the Hodge diamond.  On the other
hand,  the ``halves'' of Laplacians, defined by the ``diagonal''
$\bullet$-projections, $\omA$, $\omB$, etc., are naturally
associated with diagonal edges between neighboring vertices.
Finally, the ``quarter'' Laplacians, defined by the ``left,''
``right,''  ``up,'' or ``down'' $\bullet$-projections $\omAB$,
$\omBD$, etc.,  are associated with the centers of squares or
faces that connect four neighboring vertices in the Hodge diagram
of \fig{hodge}. Our notation is such that $\DeA{}^{p,q}$ is the
Laplacian associated with the edge that connects to vertex $(p,q)$
from the upper left, $\DeAB{}^{p,q}$ is the Laplacian associated
with the face just above vertex $(p,q)$, etc.


The ``quarter'' or ``face-centered'' Laplacians are the elementary
building blocks. The determinant of any other ``vertex'' or
``edge'' Laplacian can be constructed by multiplication. Moreover,
on a CY manifold, the Laplacians satisfy more symmetry relations
associated with symmetries of the Hodge diamond. They are induced
by  complex conjugation, Hodge $*$-conjugation, and multiplication
by the holomorphic 3-form $\Om$. For a 3-fold, these relations
leave only 3 independent faces among the total of 9 in the Hodge
diamond diagram: the 4 corner faces, the 4 edge-centered faces,
and the central face. The ``face'' Laplacians will be denoted,
respectively, as  $\Delta_A$, $\Delta_B$, and $\Delta_C$, and
their determinants as $A,B,$ and $C$.  The 9 squares
in~\fig{hodge} have been labeled accordingly.

From the Hodge decompositions, we have
    \eq{ \det' \De_{00} = \det' \De_{33} = \det' \De_{30} = \det' \De_{03} = \det\,\Delta_ A=A,}
     \eq {\alid{\det' \De_{10} =& \det' \De_{01} = \det' \De_{32} = \det'
     \De_{23}=
     \\=\det'\De_{20}=&\det'\De_{02}=\det'\De_{13}=\det'\De_{31}= \det\Delta_A  \det
    \De_B=AB,\\}}
and finally
    \eq{ \det' \De_{11} =  \det' \De_{21} = \det' \De_{12} =
    \det' \De_{22}  =\det \De_A (\det \De_B)^2 \det \De_C=AB^2C.}
The Ray-Singer torsions $I_p$ with values in the holomorphic
vector bundle $\Omega^{p,0}$ were defined in the introduction and
can be written in terms of $A,B,C$:
    \eq{ \label{ABCrel} \alid{& I_0 = I_3=(A^2 B^{-1})^{1/2} =A B^{-1/2} \\
    &I_1 =
    [(A^2 B^{-1}) (B^{2} C^{-1})]^{1/2} = A B^{1/2} C^{-1/2}.}}
    The $B$-model partition function can be similarly written
    \eq{Z_{B,1-loop}  = I_1/I_0^3 = (A^{-4} B^4 C^{-1})^{1/2}. }
Each face determinant appears to the plus or minus $1/2$ power, in
a ``chessboard'' fashion, as shown later in~\fig{final}.

After recognizing the structure and decomposition of the relevant
differential operators, one may proceed to the computation of of
the partition function for the quadratic part of the Hitchin
action,
    \eq {S_{cl} = \int_X \d b_{11} \wedge \dbar b_{11}.}
The BV complex has the following structure. We start with physical
fields $b_{11}$ and introduce ghosts $b_{10}$ and $b_{01}$, and
fields $b_{00}$ of ghost number two (``ghosts for ghosts''). The
reality conditions are obvious; $b_{pq}$ is complex conjugate to
$b_{qp}$. The statistics of $b_{pq}$ for $p,q\leq 1$ are
$(-1)^{p+q}$.  Their BRST transformations are
\eq{\label{Qcomplex1} \alid{&Q b_{11} = \dbar b_{10} + \d
b_{01},\\ &
 Q b_{10} = \d b_{00},    \quad   Q b_{01} = \dbar b_{00}.\\}}
Now we introduce antifields, which are conjugate variables of
opposite statistics.  The conjugate of a form $b_{pq}$ is a form
$b_{3-p,3-q}$, so the antifields are fields $b_{pq}$, $p,q\geq 2$,
with statistics $(-1)^{p+q+1}$.  The antibrackets are defined by
the odd symplectic structure on the phase space of fields $\la f,g
\ra =\int f \wedge g = (f ,*g)$, where $(~,~)$ denotes the metric
in the Hilbert space of $\Om^{p,q}$ forms.

Now we must write down a master action $S$, which must reduce to
$S_{cl}$ when the antifields are zero, generate the BRST
transformations \rf{Qcomplex1} when acting on the fields, and obey
$\{S,S\}=0$. In the present problem, we can take $S=S_{cl}+S'$,
where
    \eq{S'=\sum_i \la \Phi_i^*,Q\Phi_i \ra.}
Here $\Phi_i$ ranges over all fields, namely the $b_{pq}$ of $p,q\leq 1$, and
$\Phi_i^*$ are the corresponding antifields.  To prove that
$\{S,S\}=0$, we note that (i) $\{S_{cl},S_{cl}\}=0$ since $S_{cl}$
is independent of antifields; (ii) $\{S_{cl},S'\}=0$, since when
acting on fields, $S'$ generates the transformation
\rf{Qcomplex1}, under which $S_{cl}$ is invariant; (iii)
$\{S',S'\}=2\sum_i\int\Phi_i^*Q^2\Phi_i,$ which vanishes, since
$Q^2=0$ on fields.

Having found the master action, we can also find the BRST
transformations of antifields, as these are generated by
$Q\Phi_i^*=\{S,\Phi_i^*\}$: \eq{\alid{
&Qb_{22}=\partial\bar\partial b_{11} \\ & Qb_{32}=\partial b_{22},
~ Qb_{23}=\bar\partial b_{22}.\\}}

Now we must choose a  Lagrangian submanifold (again for simplicity
neglecting the Hodge cohomology groups). As in our practice
example, we define the Lagrangian submanifold to be the
orthocomplement of the space of gauge variations. This can be
defined using the projection operators that we have already
constructed, and removes the kernels of all kinetic operators that
appear in the master action.

In detail, the Lagrangian submanifold is defined by saying that
for some forms $\alpha_{pq}$ of appropriate degrees, with
$\alpha_{qp}=\bar\alpha_{pq}$, we have \eq{\alid{ b_{11} & =
\partial^*\bar\partial^* \alpha_{22} \\
b_{10}&=\partial^*\alpha_{20}+i\partial\alpha_{00} \\
b_{01}& = \bar\partial^*\alpha_{02}-i\bar\partial\alpha_{00}.\\}}
No condition is placed on $b_{00}$.  Once we have selected these
conditions on the fields, the corresponding conditions on the
antifields are uniquely determined; the antifields must be
constrained to obey \eq{\sum_i\int \Phi_i^*\wedge \Phi_i=0,} and
no other conditions.  For example, since we have placed no
constraint on $b_{00}$, we must impose \eq{b_{33}=0.} The ``dual''
of imposing $b_{11}=\partial^*\bar\partial^*\alpha_{22}$ is to
impose \eq{b_{22}
=\partial^*\alpha_{32}+\bar\partial^*\alpha_{23}.} Finally, having
constrained $b_{10}$ and $b_{10}$ as above, to get a Lagrangian
submanifold, we must place the ``conjugate'' conditions on
$b_{32}$ and $b_{23}$: \eq{\label{pl}\alid{ b_{32}&=\partial^* \alpha_{33} \\
                                  b_{23}& =\bar\partial^*
                                  \alpha_{33}.\\}}
In verifying that these conditions define a Lagrangian submanifold
of the space of fields, we must show that they imply that
$\int(b_{23}\wedge b_{10}+b_{32}\wedge b_{01})=0$.  To show this,
one substitutes in \rf{pl}, integrates by parts, and uses that the
Laplacians $\bar\partial^*\bar\partial$ and $\partial^*\partial$
are equal.

The purpose of this choice of a Lagrangian submanifold is that it
makes the evaluation of the path integral straightforward. For
example, the only term in $S$ that contains the original physical
field $b_{11}$ is the original classical action $S_{cl}=\int
b_{11} \wedge \dbar \d  b_{11}$.  Our Lagrangian submanifold
simply projects $b_{11}$ onto the space $\omCD^{11}$, associated
with the square ``below'' the 11 vertex of the Hodge diagram; this
is the central square. The projected kinetic operator is hence
simply $\DeCD^{11}$, whose determinant is what we have called $C$.
As $b_{11}$ is a real bosonic field, its path integral is
therefore $C^{-1/2}$.


Next, we look at the part of the path integral that involves
$b_{00}$.  The relevant terms in $S$ are $\int\left(b_{23}\wedge\d
b_{00}+b_{32}\wedge \dbar b_{00}\right)$.  We can write this as
$(b_{23}+b_{32},Wb_{00})=\int (b_{23}+b_{32})\wedge * Wb_{00}$,
where the operator $W$ is \eq{ W = \Pi
*   (\partial+\bar\partial).} Here $\partial+\bar\partial$ maps
$\Omega^{00}$ to $\Omega^{10}\oplus\Omega^{01}$, the Hodge $*$
maps this to $\Omega^{23}\oplus \Omega^{32}$, and finally $\Pi$
projects to the intersection of the Lagrangian submanifold with
$\Omega^{23}\oplus \Omega^{32}$.  The result of the path integral
over $b_{00}$, $b_{23}$, and $b_{32}$ is therefore $\det^{-1}W$.
Here $\det \,W$ is raised to a negative power because these fields
are bosonic, and the power is $-1$ because one has a pair of
independent fields, namely $b_{00}$ and a linear combination of
$b_{23}$ and $b_{32}$.  On the other hand, $\det W = (\det
W^*W)^{1/2}$, and $W^*W$ is the Laplacian $\De_A=\DeCD^{00}$
associated with the top square of the Hodge diamond. We have
called the determinant of this operator $A$. So finally, the
integral over these fields gives $A^{-1/2}$.

Finally, we consider the path integral over $b_{10}$, $b_{01}$,
and $b_{22}$.  The relevant action after integrating by parts is
$\int (b_{01}\partial+b_{10}\bar\partial)b_{22}=
((b_{10}+b_{01}),W'b_{22})$, where rather as before $W'=\Pi *
(\partial+\bar\partial)$.  $W'$ is a map from $\Omega^{22}$ to the
intersection of the Lagrangian submanifold with $\Omega^{10}\oplus
\Omega^{01}$.  The fields are now fermionic, so the result of the
path integral is $\det W'=(\det W'{}^*W')^{1/2}$.  In this case,
because the intersection of the Lagrangian submanifold with
$\Omega^{10}\oplus\Omega^{01}$ is relatively complicated, the
description of $W'{}^*W'$ is more complicated than what we
encountered in the last paragraph.  $W'$ maps $\Omega^{22}$ to the
subspace of $\Omega^{10}\oplus \Omega^{01}$ associated with the
top three squares of the Hodge diamond -- the top square with
vertex $00$ labeled $A$ in~\fig{hodge}, and the two adjacent squares
labeled $B$. The spectrum of $W'{}^*W'$ is therefore that of
$\DeCD^{00}\oplus \DeCD^{10}\oplus \DeCD^{01}$ (this operator is
in fact $W'W'{}^*$), and the determinant of $W'{}^*W'$ is $AB^2$.
Hence the path integral over these fields is $A^{1/2}B$.

Overall, then, the one-loop path integral of the minimal Hitchin
model is $C^{-1/2}B$.  Referring back to~\rf{ABCrel}, we see that
this is precisely \eq{\label{H1loop} Z_{H,1-loop}={I_1\over I_0},}
a product of Ray-Singer analytic torsions, but not the partition
function of the $B$-model, which is
 \eq{ \label{ZB1loop} Z_{B,1-loop} = I_0^0  I_1^{-1} I_2^{2} I_3^{-3} =
 I_1/I_0^3.
    }
In terms of the Hodge diagram, $Z_{H,1-loop}$ involves only the
middle diagonal of faces, while  $Z_{B,1-loop}$ involves also the
side diagonals on the Hodge diagram.  Two factors of $I_0$ are
missing in~\rf{H1loop} compared with~\rf{ZB1loop}.

\EPSFIGURE{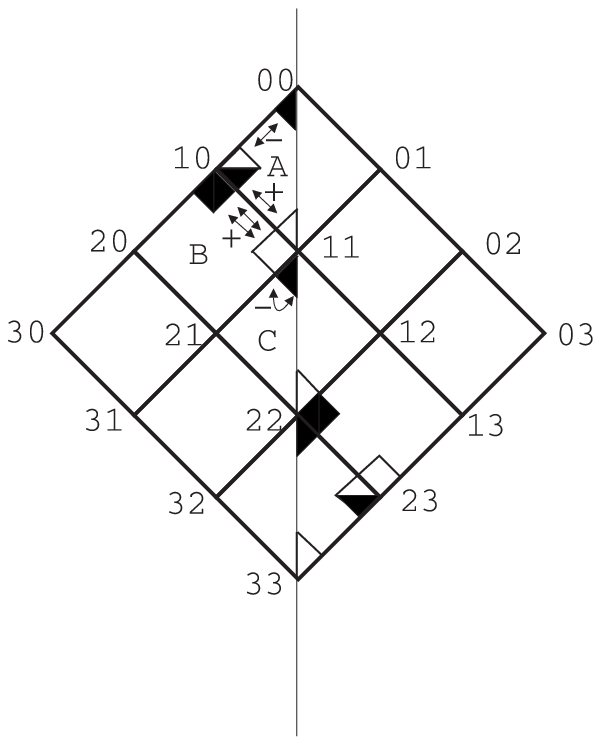} { \label{minimal} The projection to the
chosen Lagrangian submanifold is shown by the black filling. The
arrows between vertices represent quadratic terms in the master
action $S$ for $\int_X \d \dbar b_{11} \wedge b_{11}$. The Hodge
operator $*$ works as a central reflection of the picture. The
fields are located above the center of the diamond, antifields are
below. }

When we consider the generalized Hitchin action, such a detailed
description of the evaluation of the path integral would be rather
messy. Fig. \ref{minimal} is designed to give a short cut. This
figure is meant to give a convenient way of describing the
Lagrangian submanifold, the differential maps between different
Hilbert spaces and the resulting powers of determinants.  The
fields $b_{00}$, $b_{10}$, and $b_{11}$ take values in
$\Omega^{00}$, $\Omega^{10}$, and $\Omega^{11}$, respectively, and
live at the appropriate vertices in the Hodge diamond. Each term
in the Lagrangian contains one of these three fields, or their
complex conjugates. (To keep the figure simple, the complex
conjugate fields have been omitted.) The spaces $\Omega^{00}$,
$\Omega^{10}$, and $\Omega^{11}$ have Hodge decompositions, as
described above and summarized in \fig{squares} and \fig{real}.
The restriction to the Lagrangian submanifold projects $b_{00}$,
$b_{10}$, and $b_{11}$ onto subspaces of $\Omega^{00}$, etc. The
components of these fields associated with certain little squares
and little triangles are set to zero by the projection, and other
components are not. In~\fig{minimal}, we have filled in black the
little squares and triangles corresponding to nonzero components
of the fields.

The Lagrangian projection for the antifields is obtained by
rotating the figure through 180 degrees around its center and
exchanging white and black.  This exchange of colors is equivalent
to the condition that the projection is Lagrangian.  The fact that
the Lagrangian submanifold is the projection to the
orthocomplement of the image of $Q$ is expressed in the picture in
the fact that $Q$ always maps a black region in $\Omega^{pq}$ to a
white region in $\Omega^{p+1,q}$ or $\Omega^{p,q+1}$.  This
ensures that the projected action is nondegenerate.

To find the powers of $A$, $B$, and $C$ in the partition function,
we just look at how the squares labelled $A$, $B$, and $C$ in the
figure are shaded.  In the square $C$, for example, the only
shaded region is a single triangle, or half of a small square,
associated with the real bosonic field $b_{11}$.  The resulting power
of $C$ is therefore $C^{-1/2}$, where the minus sign reflects the
fact that $b_{11}$ is bosonic, and the exponent $1/2$ reflects the
fact that what is shaded is a triangle or $1/2$ of a small square.
In the case of $A$, the shaded regions are two small triangles,
one associated with a bosonic field $b_{00}$ and one with a
fermionic field $b_{10}$.  They contribute $A^{-1/2}$ and
$A^{1/2}$, respectively, so $A$ does not appear in the partition
function of this model.  Finally, for $B$, we have a shaded square
associated with the fermionic variable $b_{10}$, so the
corresponding factor in the partition function is $B$.  Putting it
all together, the partition function of this model is again
$BC^{-1/2}$.

One point in this description really needs a fuller explanation.
The field $b_{11}$, unlike $b_{00}$ and $b_{01}$, has a second
order action.  Other things equal, this would double the exponent
in its determinant.  But this field is self-conjugate, the action
being $\int b_{11}\partial\bar\partial b_{11}$, while $b_{00}$ and
$b_{01}$ are conjugate to other fields $b_{pq}$ with $p$ and $q$
greater than 1.  A second order action for a self-conjugate field
or a first order action for a pair of fields each give a
determinant of the appropriate Laplacian to the plus or minus 1/2
power.  So we get again
    \eq{ Z_{H,1-loop} = \lb \det^{-1} \De_C \det^{2} \De_B \rb^{1/2}.}

Let us explain this formula in a spirit of the formal evaluation
of the path integral as in~\rf{strpath}.  Here we will be more
complete and include the cohomology groups.  First we give a
formal expression for the path integral over the original physical
field $b_{11}$:
    \eq{Z_{H,1-loop} = \frac {1}{{\rm vol}(G)}
    \det^{-1/2} \De_C \vol^{1/2}(H^{11}) \vol (\omAC^{11})
    \vol^{1/2} (\omAB{}^{11})}
Here the factor $\det^{-1/2}\De_C$ comes from the Gaussian
integral over the part of $b_{11}$ that is orthogonal to the gauge
variations, and $\vol(G)$ is the volume of the gauge group.  We
also need a factor  to account for the volume of the space of
$b_{11}$ fields that {\it do} arise as
$\partial\lambda_{01}+\bar\partial\lambda_{10}$. The space of such
$b_{11}$ fields is shown in~\fig{minimal} by a small white square and a
small white triangle at the 11 vertex.  The white square denotes
$b_{11}$ fields that are of the form $\bar\partial\lambda_{10}$
plus complex conjugate, but orthogonal to fields of the form
$i\bar\partial\partial \lambda$ for real zero-forms $\lambda$,
while the white triangle denotes fields $b_{11}$ that are of the
form $i\bar\partial\partial\lambda$, for real $\lambda$. So the
volume of $b_{11}$ fields that are gauge variations is
  $\vol (\omAC^{11})  \vol^{1/2} (\omAB{}^{11})$. Finally, we have
included  in the partition function a
    factor of $ \vol^{1/2}(H^{11})$, the volume of the real
    subspace of $H^{11}$, to account for the volume of the
    harmonic part of $b_{11}$.

 We can compute the volumes of spaces of forms as we did in the real case.  We
have to be careful on one point: if $w:V\to W$ is an invertible
map between complex vector spaces, then the ratio of volumes of
$W$ and $V$ is $\det (w^*w)$ (while in the real case it is
$\det^{1/2}(w^*w)$). Taking $w$ to be the $\bar\partial$ operator
from $\omCD{}^{10}$ to $\omAC{}^{11}$, or from  $\omBD^{10}$ to
$\omAB{}^{11}$, we get
   \eq{  \vol(\omAC^{11}) = \det \De_B     \vol(\omCD^{10}) }
and \eq{
        \vol^{1/2} (\omAB{}^{11}) = \det^{1/2} \De_A  \vol^{1/2}{(\omBD^{10})}
        .        }
Also, as the volume of $\Omega^{10}$ is the product of the volumes
of $\omCD^{10}$,  $\omBD^{10}$, and $H^{10}$, we have
    \eq{  \vol(\omCD^{10}) \vol^{1/2}{(\omBD^{10}) } = \f {\vol (\Om^{10})} {
    \vol^{1/2}{(\omBD^{10})} \vol{(H^{10}) }}.}
Using the $\partial$ operator from $\Omega^{00}$ to the part of
$\Omega^{10}$ represented by the upper small triangle, one has
similarly
    \eq{ \vol^{1/2}( \omBD^{10}) = \det^{1/2} \De_A \vol^{1/2} (\omCD^{00}).}
Also, $\vol(\Om^{00})=\vol(\omCD^{00})\cdot \vol (H^{00})$.
 Bringing all this together, one has
    \eq{ Z_{H,1-loop} = \f{1}{\vol(G)}
    \f {\vol^{1/2}(H^{11})\vol (H^0)}{\vol(H^1)} \det^{-1/2} \De_C \det \De_B \f
    {\vol(\Om^{1})}{\vol(\Om^{0})},}
where $\Om^{0}$ is the space of \emph{real} 0-forms and $\Om^{1}$
is the space of \emph{real} 1-forms. Also, $H^0$ and $H^1$ are the
\emph{real} de Rham cohomology groups (so for example
$\vol(H^0)=\vol(H^{00})^{1/2}$).  The easiest way to keep track of
these formulas is to use the picture of \fig{minimal}.

We suppose that $\vol(G)=
    {\vol(\Om^{1})}/{\vol(\Om^{0})} $, the intuitive idea being
    that the gauge group consists of gauge transformations of the
    physical fields and the ghosts by one-forms and zero-forms,
    respectively.
 So \eq{Z_{H,1-loop}=\f {\vol(H_\BR^{11})\vol
(H^0)}{\vol(H^1)} \det^{-1/2} \De_C \det \De_B,} the same result
that we got with the BV formalism, except that now we have
included the spaces of zero modes.  Here $H^{11}_\BR$ is the real
subspace of $H^{11}$, so $\vol(H^{11}_\BR)=(\vol(H^{11}))^{1/2}$.


\underline{The quadratic term of the extended Hitchin functional}

\EPSFIGURE{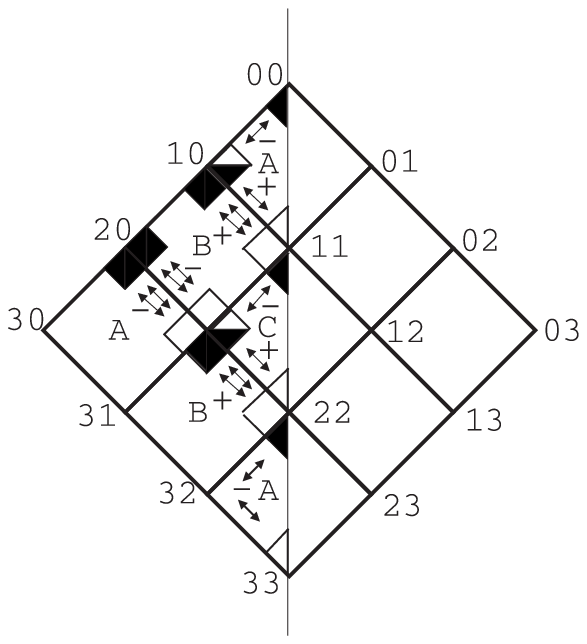}{ \label{extended} This picture explains
the computation of determinants for the term $\int_X a\d \dbar
b_{22} $. The notation is the same as in~\fig{minimal}, except
that here only $\tilde a$ and the $b_{pq}$ are drawn. The picture
and the Lagrangian subspace for the antifields $\tilde b_{pq}$ and
the field $a$ is obtained precisely by 180 degree rotation and
flipping black and white colors.
  The picture encodes the
Lagrangian submanifold, and the arrows that represent mappings
between various Hilbert spaces of forms. The resulting
determinants  all cancel \emph{except} those associated with the
three squares on the lower left edge. The result is $A^{-1} B
A^{-1}$. }

Since the minimal Hitchin model appears to disagree with the
$B$-model at one-loop order, we consider now the extended Hitchin
functional $H(\phi)$. As we explained in section 2, this
functional is defined for a polyform $\phi = \rho_1 + \rho_3 +
\rho_3$. It is constructed by Hitchin~\cite{Hitchin:2} in a
similar spirit to the minimal Hitchin functional $H(\rho)$ and the
action is also given by a square root of a  $4^{th}$ order
polynomial in $\phi$. The critical points of $H(\phi)$ determine a
\emph{generalized CY structure}\cite{GCS,Hitchin:2} on $X$. If
$b_1(X)$ is trivial, then this generalized CY structure is
equivalent to a standard CY structure. The quadratic expansion of
the extended Hitchin functional is also given as $ \int_X \de \phi
\wedge J \de \phi$, where $J$ is a complex structure on a space of
odd polyforms $\phi$ introduced by Hitchin. With respect to the
Hodge decomposition of forms, it has $+i$ eigenvalues on the
spaces $\Om^{1,0}, \Om^{3,0}, \Om^{2,1}, \Om^{3,2}$. The quadratic
expansion of the extended Hitchin functional near a critical point
has the following form
    \eq {S_{cl} = \int_X \left( b_{11}\wedge \d \dbar b_{11} + b_{00} \wedge \d \dbar b_{22}\right).}

We need to compute the partition function associated with the
second term.  It will be helpful to change notation slightly and
write $a$ for $b_{00}$, enabling us to reserve the name $b_{00}$
for a ghost field that will arise presently.  The classical action
is thus \eq{S_{cl}=\int a\wedge \d\dbar b_{22}.} The gauge
invariance is of course $\delta b_{22}=\partial
b_{12}+\bar\partial b_{21}$. To introduce the BV complex, we begin
by introducing the whole collection of ghost fields $b_{pq}$,
$p,q\leq 2$, with \eq {\label{Q_complex2} \alid{
                &Q b_{22} = \dbar b_{21} + \d b_{12} \\
                &Q b_{21} = \dbar b_{20} + \d b_{11}, \quad  Q b_{12} = \dbar b_{11} + \d b_{02}
                \\
                &Q b_{20} = \d b_{10}, \quad Q b_{11} = \dbar b_{10} + \d
                b_{01},
                \quad Q b_{02} = \dbar b_{01} \\
                &Q b_{10} = \d b_{00},   \quad Q b_{01} = \dbar b_{00} \\
                }}
Here $b_{pq}$ has ghost number $4-p-q$ and statistics
$(-1)^{p+q}$.  Next, we introduce antifields.  The antifield for
$a$ is a fermionic $(3,3)$-form $\tilde a$.  The antifields of
$b_{pq}$ are fields $\tilde b_{3-p,3-q}$, of statistics
$(-1)^{p+q+1}$ and ghost number $p+q-5$.  The master action is
\eq{S=S_{cl}+\sum_{p,q}\int \tilde b_{3-p,3-q}\wedge Qb_{pq}.} It
obeys all the necessary conditions by virtue of the same arguments
that we gave in quantizing the original Hitchin Lagrangian.  Note
that to go from $S_{cl}$ to $S$, we do not add any terms depending
on $a$ and $\tilde a$, since $a$ is gauge invariant and does not
appear on the right hand side of  any transformation laws in
\rf{Q_complex2}.  The transformation laws of antifields under $Q$
can be obtained as $Q\tilde b=\{\tilde b,S\}$, $Q\tilde a=\{\tilde
a,S\}$. We will not write them in detail.

The next step is to introduce a Lagrangian submanifold.  As in the
previous discussion, we do this by projecting each field onto the
subspace orthogonal to its variation under a gauge transformation.
The evaluation of the path integral is then straightforward, just
as we have seen earlier, with each set of fields contributing
determinants associated with certain faces of the Hodge diagram.
However, the details are lengthy, because there are so many
fields.  Hence, it is extremely helpful to take a short cut using
a figure analogous to~\fig{minimal}.

In the~\fig{extended}, we have sketched the various fields and
their Hodge decomposition after projection to the Lagrangian
submanifold. Each square in the Hodge diamond has an accompanying
determinant, which will appear in the partition function with some
exponent that we want to compute. The exponent for each square is
obtained by counting black regions in that square derived from
bosonic or fermionic fields.  Many squares do not contribute.  For
example, the top square in the Hodge diagram has two black
triangles, one associated with a bosonic field and one with a
fermionic field, so the net exponent is zero.  The neighboring
square with top corner labeled 10 contains two small black
squares, one derived from a bosonic field and one from a fermionic
field, so the overall exponent is again zero.  The center square
has two black triangles, associated with fields of opposite
statistics, again giving no net contribution.

Finally, the squares that do contribute are the three squares on
the lower left of the Hodge diamond, with left vertex 30, 31, or
32. The square with left vertex 30 contains a small black square
derived from a bosonic field, so it contributes a factor $A^{-1}$
to the path integral. The square with left vertex 31 contains a
small black square derived from a fermionic field, so it
contributes a factor $B$. Finally, the bottom square in the Hodge
diagram contains a single black triangle associated with the
original bosonic field $b_{22}$. Though a triangle usually gives
us an exponent of $\pm 1/2$, in this case $b_{22}$ has a separate
conjugate variable, with which it appears in the underlying second
order classical action $\int a\partial\bar\partial b_{22}$.  As
there are two fields and a second order action, the factor
contributed to the path integral is $A^{-1}$.  Putting all this
together, the extra factor in the partition function of the
extended Hitchin model compared to that of the minimal Hitchin
model is $BA^{-2}$.

The result for the partition function of the extended Hitchin
model is thus\ft{When the cohomology groups are nonzero, they
contribute to $Z_{HE}$ an additional factor
    \eq{  \f {\vol^{1/2}(H^{11})\vol
(H^0)}{\vol(H^1)}  \vol^{1/2} (H^{00}) \cdot \f {\vol^{1/2}
(H^{22}) \vol (H^{20}) \vol^{1/2} (H^{11}) \vol^{1/2} (H^{00})}
{\vol (H^{21}) \vol (H^{10}) } } }
    \eq {Z_{HE,1-loop} = BA^{-2} Z_{H,1-loop} = A^{-2}B^2C^{-1/2}, }
    or equivalently, in terms of torsions,
    \eq { Z_{HE,1-loop}  =\frac{ I_1}{ I_0^3} = \prod_{p=0}^{3} I_p^{ (-1)^{p} p}.}
This agrees with the one-loop partition function of the $B$-model,
according to \rf{bactors}.

We summarize the final result in \fig{final}.

\EPSFIGURE{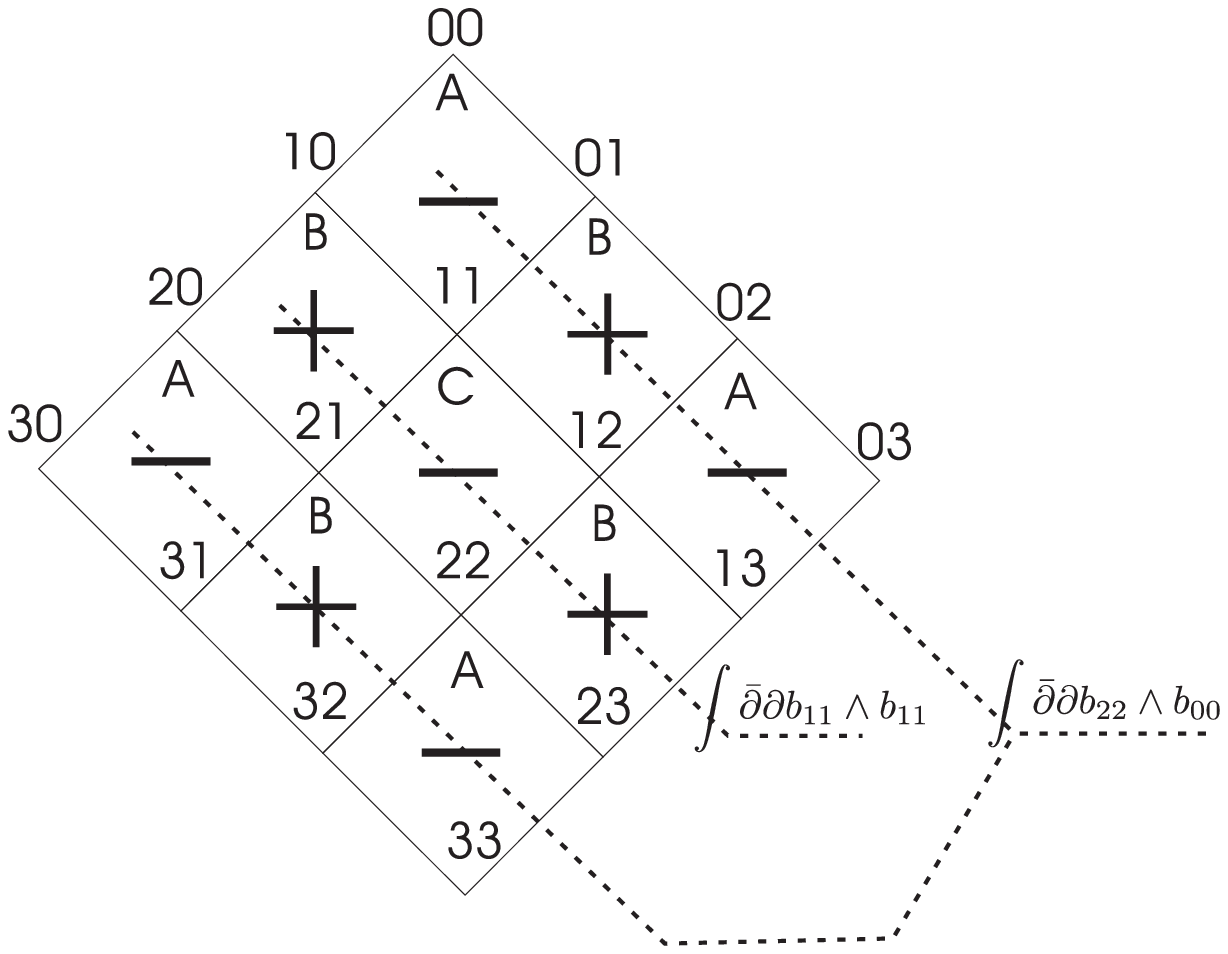}{The ``chessboard'' structure for the powers
of ``face'' determinants is shown for the
$Z_{HE,1-loop}=Z_{B,1-loop}$ \label{final} }

\bigskip

\section{Gravitational Anomaly}

Now we want to investigate the dependence of the quantum Hitchin
and extended Hitchin models on the background metric that is used
for quantization.

This metric is absent in the initial or classical description of
the theory, but it has been used to define the Lagrangian
submanifold used for quantization and the regularized determinants
of Laplace operators.  If the resulting partition function depends
on the metric, this is an anomaly that seriously affects the
physical interpretation of the model.

The dependence of Ray-Singer torsion on the Kahler metric has been
determined in \cite{MR929146,MR929147,MR931666}.  The general
formula applies to the torsion of the $\bar\partial$ operator with
values in any holomorphic vector bundle $V$ on a complex manifold
$X$ of complex dimension $n$; for us, the case of interest is
$V=\Omega^{p,0}$. The result has been worked out in more detail,
for $p=0$, in \cite{MR1714484}. In the case of complex dimension
one, as was originally elucidated in \cite{Belavin:1986cy}, the
dependence of the determinants of Laplacians on the metric reduces
to the usual conformal anomaly that appears in Polyakov's
quantization of the bosonic string \cite{Polyakov:1981rd}.

We will describe the variation of the torsion with the metric for
a general holomorphic vector bundle $V$, and then specialize to
the case that $V$ is some $\Omega^{p,0}$.  (One reason to begin
with the generalization is that it is actually relevant to the
$B$-model in the presence of $D$-branes.)  We write $g$ for a
Kahler metric on $X$ and $h$ for a hermitian metric on $V$.  We
want to describe the change in the torsion under general
deformations $g\to g+t\delta g$, $h\to h+t\delta h$, with $t$ a
small parameter and $\delta g$ and $\delta h$ arbitrary small
variations (preserving of course the Kahler and hermitian
conditions). We write $R$ for the Riemann curvature tensor derived
from $g$ and $F$ for the curvature of the connection derived from
$h$.

Before trying to present the formula for the variation of the
torsion, it helps to develop some notation.  As an example, the
first Chern class of a holomorphic vector bundle $V$ corresponds
to the differential form $c_1=\tr(\f i {2\pi} F )$.  The curvature
$F$ is, of course, a $(1,1)$-form on $X$ with values on $\End(V)$.
Now we will consider an expression
    \eq{ c_1(V(t)) = \tr \f {1} {2\pi} \lb iF+t h^{-1} \de h\rb.
    }
Here $h^{-1}\delta h$ is a $(0,0)$-form on $X$ with values in
endomorphisms of $V$, so after taking the trace we get the sum of
a $(1,1)$-form on $X$ and a $(0,0)$-form.  We will think of this
combination as a first Chern class in an extended sense.

The $t$ derivative of it is
    \eq{ \dot c_1(V(t)) = \f{1} {2\pi} \tr h^{-1}\delta h.}
As we see, though the ordinary $c_1$ is a $(1,1)$-form, the
$t$-derivative of the extended $c_1$ is a $(0,0)$-form.

Similarly, for a higher Chern class $c_k(V)$, which is defined as
a certain homogeneous $k^{th}$ order function of $\f i {2\pi} F$,
we define an extended, $t$-dependent version $c_k(V(t))$ by
replacing $iF$ with $iF+th^{-1}\delta h$. Then     $\dot
c_k(V(0))$ is a $(k-1,k-1)$-form, though the ordinary $c_k(t)$ is
of course a $(k,k)$-form.

For a general characteristic class like the Todd class $\td(TX)$
or the Chern character $\ch(V)$, we write $\td(TX)_k$ or
$\ch(V)_k$ for the components of degree $(k,k)$.  These functions
are expressed as homogeneous polynomials of degree $k$ in the
curvatures $\f i {2\pi} R$ and $\f i {2\pi} F$, so we can define
$t$-dependent generalizations by replacing $iR$ by $iR+t
g^{-1}\delta g$ and $iF$ by $iF+th^{-1}\delta h$.  Clearly, then,
    $\left.\frac{\partial}{\partial t}\td(TX)_k\right|_{t=0}$
is of degree $(k-1,k-1)$, and likewise if we replace $\td(X)$ by
$\ch(V)$ or a product of similar expressions. For a general
holomorphic vector bundle $V$, one defines
    \eq{ c(V) \equiv c \lb \f i {2 \pi} F\rb = \det \lb 1 + \f i {2\pi} F\rb =
    \prod_{i=1}^r (1+x_i)= \sum_{k=0}^{r} c_k(V),
    }
where $x_i$ are the formal eigenvalues of $\f {i} {2\pi} F$ and
$r=\rk(V)$.  So
    \eq{ c_k (V)  = \sum_{i_1<i_2<\dots i_k} x_{i_1}x_{i_2}. \dots x_{i_k}}
Then
    \eq {\td(V) = \prod_{i=1}^{r} \f {x_i} {1 - e^{-x_i}}, \quad  \ch(V) = \sum_{i=1}^r e^{x_i}}

 According to  Theorem 1.22 in~\cite{MR931666}, \ft{This theorem still holds with nonzero $H^{pq}$, as long
 as the definition of the analytic torsion is appropriately extended to include the volumes
 of the cohomology groups.} the variation of the torsion under a change in
 Kahler metric  is
 given by
    \eq { \f {1} {2\pi} \label{varformula}\left. \f {\d} {\d t}\right|_{t=0}
     \log I(X,V,g,h) = \f 1 2 \int_X \f {\d} {\d t}
    \left[ \td \f {1} {2\pi}( iR+ t g^{-1} \de g) \ch \f {1} {2\pi}(iF+ t h^{-1} \de h)
\right]_{n+1}.
    }
For the closed string $B$-model, we specialize to the case that
$V$ is a sum of bundles $\Omega^{p,0}$, $F$ coincides with $R$
(written of course in the appropriate representation $\Om^{p,0} =
\bigw^p {T^*X})$ and $h$ is determined by $g$.

For the  $B$-model or equivalently the extended Hitchin model at
one-loop, one has
    \eq{ \label{F1var} F_1^B = - \sum_{p=0}^{n} (-1)^{p} p \,\log I(X,\Om^{p,0}),}
where we do not indicate the dependence on the metric explicitly.
The anomaly of the ordinary Hitchin model can of course be studied
in a similar way, but apparently does not lead to such simple
formulas. We will omit this case.

To compute the variation of $F_1^B$ using ~\rf{varformula}, it is
first convenient to algebraically simplify the characteristic
class entering into the variation of $F_1$.  This class is
    \eq{ P =\sum_{p=0,\dots,n} p(-1)^p \td(TX) \ch(\wedge^p T^*X),}
where $TX$ is the holomorphic tangent bundle of $X$, and $T^*X$ is
its dual. A simple expression for $P$ was obtained
in~\cite{Bershadsky:1993cx}, where it arose for  reasons closely
related to our present discussion.

 Let $x_i$ be the Chern roots of $TX$ (the formal eigenvalues of $\f {i} {2\pi} R$, where $R$ is
 the curvature of $TX$). The Chern roots of $\wedge^p T^*X$ are
$-x_{i_1}-x_{i_2}-\dots -x_{i_p}$, $i_1< i_2<\dots <i_p$.  So the
Chern character of $\wedge^p T^*X$ is $\sum_{i_1<\cdots < i_p}
    e^{-x_{i_1}-x_{i_2}-\dots-x_{i_p}}$.

We have then
    \eq{ \alid{
    \sum_{p=0}^n(-1)^pp \ch(\wedge^pT^*X) &= \sum_{p=0,\dots,n} p (-1)^p  \sum_{i_1<\cdots < i_p}
    e^{-x_{i_1}-x_{i_2}-\dots-x_{i_p}} \\
    &=
    \d_z \prod_{i=1}^{n} (1 - z e^{-x_i}) |_{z=1} = -
     \sum_{j=1,\dots,n} e^{-x_j} \prod_{\substack{i=1,\dots,n \\i \neq j}}
    (1-e^{-x_i}).\\}}
Hence, after multiplying by the Todd class to get $P$, we have
\eq{\alid{P &=
     - \prod_{i=1}^n x_i \sum_{j=1}^n \f
    {e^{-x_j}} {1 - e^{-x_j}} = \sum_{j=1}^{n} \f {x_j} {1-e^{x_j}}
    \prod_{\substack{i=1,\dots,n \\i \neq j}} x_i  = - \sum_{j=1}^n
    \left(1 - \f {x_j} 2 + \f {x_j^2} {12} + \dots\right) \prod_{\substack{i=1,\dots,n \\i \neq j}}
    x_i \\ &= -\left(c_{n-1} - \fr n 2 c_n + \fr 1 {12} c_n c_1 + \dots
    \right)
    }
    }
The component of degree $n+1$ is equal simply to $-\fr 1 {12} c_n
c_1$.

Plugging this into the variational formula~\rf{varformula}, one
has
    \eq{ \f {1} {2\pi}\f {\d} {\d t} F_1(g+t\de g) = - \f 1 {24}  \int_X \d_t (c_n c_1).}
As we have already discussed, the $t$ derivative  of $c_1(TX(t))$ is given by
    \eq{ \dot c_1 = \f {1} {2\pi} \tr g^{-1} \de g}
The derivative  $c_n$ can be similarly expressed in terms of $\tr
R^k g^{-1} \de g$, but one does not need the explicit expressions
to see that generally, for an arbitrary metric $g$ and an
arbitrary variation $\de g$,
 the total variation $\int_X  \d_t (c_n c_1)$ is nonzero and more
 seriously, is not a function only of the Kahler class.

Consider a general variation of the metric that does not change
the Kahler class.  Thus, $\delta g=\partial\bar\partial u$, for
some function $u$, and $\tr g^{-1}\delta g=\Delta u$, $\Delta $
being the Laplacian.  If the complex dimension of $X$ is $n=1$,
the variation of $F_1$ involves
    \eq{\int_X \d_t(c_1^2)=2 \f{1} {2\pi} \int_X c_1\Delta u.}
In general $c_1$ is a completely arbitrary $(1,1)$ curvature form
whose integral is zero, and $u$ is an arbitrary function; this
expression certainly does not vanish in general. Similarly, for
complex dimension three, the anomaly does not vanish in general.
 For example, if one takes $X$ to be the product of a K3 surface and an elliptic curve $E$, with a fixed
 Ricci-flat metric on K3 and a varying Kahler metric on $E$, then the anomaly computation
 reduces to what we have just described.
 So the $B$-model or extended Hitchin model is anomalous
 at one-loop order.

We really do not know how to interpret this result in general.
However, we can make one mildly comforting remark: the variation
of the one-loop $B$-model partition function with respect to the
metric is a function only of the Kahler class if evaluated for a
Ricci-flat Kahler metric $g$ and a deformation $\de g$ that
preserves the Ricci-flatness.   To show this, first note that the
Ricci-flat condition means that the $(1,1)$-form representing
$c_1$ is zero \emph{pointwise} on $X$. So if we compute at a
Ricci-flat Kahler metric, the variation of $\f {1} {2 \pi} F_1^B$
is
    \eq{ \f {1} {2\pi}\f {\d} {\d t} F_1(g+t\de g) = - \f 1 {24}  \int_X c_n \dot c_1,}
so
    \eq{ \de F^B_1 = -\f 1 {24} \int_X c_n \tr g^{-1} \de g
    }

But as we will now show, if we vary the metric in a way that
preserves the Ricci-flat condition, then $\dot c_1$ is also
simple, and the derivative of  $F_1^B$ with respect to the Kahler
moduli is only a function of  the Kahler moduli, not the complex
structure moduli. Moreover, the Kahler moduli enter only via the
volume of $X$.  This means that as long as we compute only for
Ricci-flat metrics, we can remove the anomaly, somewhat
artificially, by multiplying the one-loop partition function by a
factor that only depends on Kahler moduli, in fact, only on the
volume of $X$.

So consider a variation $\de g$ of the Kahler metric, which
preserves the Ricci-flat condition while (inevitably) changing the
Kahler class. Since the variation of the Ricci tensor
    \eq{ R_{i\bar j} = - \d_i \d_{\bar j} \log \det g}
must vanish, one has the condition on variation $\de g$:
    \eq { \d_i \d_{\bar j} \tr g^{-1} \de g = 0.}
 Hence if $X$ is compact or suitable boundary conditions are placed at infinity, then
 $ \tr g^{-1} \de g$ is constant on $X$, and hence can be
 identified with the change in the volume:
 \eq{\tr g^{-1}\de g=2\de \log {\rm Vol(X)}.}
So the formula for variation of $F_1$ simplifies to
    \eq { \de F_{1} = - \f {1} {24} \tr (g^{-1} \de g)  \int_X
    c_n= - \f 1 {12} \de \log{\rm Vol}(X) \chi(X)
    }
\emph{Hence for $Ricci$ flat metric preserving variations
\eq{\label{gurg} Z_{B,1-loop}{\rm Vol}(X)^{-\chi/12}.} is
independent of the Kahler moduli of $X$}.  We should note,
however, that in this paper we have not very well understood the
physical meaning of the zero modes that appear in quantizing the
various actions that we have considered -- especially the zero
modes for ghosts and antifields.  Since the volume of $X$ is a
function of the zero modes (of the physical fields), it may be
that the physical meaning of the volume factor in \rf{gurg} really
needs to be considered along with the physics of the zero modes.
At any rate, we recall that mathematically, when zero modes are
present, the torsion, and hence $Z_{B,1-loop}$, is not a number
but a measure on the space of zero modes.

One additional simple remark is that the anomaly suggests that one
should embed the $B$-model and the extended Hitchin model in a
target space theory which involves the Kahler metric and produces
a Ricci-flat metric in its critical points.  An obvious candidate
is the Hitchin functional in seven dimensions that leads at its
critical points to metrics of $G_2$ holonomy, specialized to
seven-manifolds of the form $X\times S^1$.  This idea is in accord
with the philosophy proposed in
\cite{Dijkgraaf:2004te,Gerasimov:2004yx,Nekrasov:2004vv,Hitchin,Hitchin:2}.



\section{Conclusion}

   The Hitchin functionals are closely related to
topological strings in target space as has been demonstrated at
tree level
~\cite{Dijkgraaf:2004te,Gerasimov:2004yx,Nekrasov:2004vv}.

The first quantum effects appear in  one-loop order. The quadratic
computation has been made in the present work using the
Batalin-Vilkovisky formalism, for which it is well adapted because
of the presence of redundant or reducible gauge transformations.
The result shows that the minimal Hitchin model does not agree at
one-loop order with the conjecture of \cite{Dijkgraaf:2004te}, but
the extended Hitchin model does.

A puzzle remains. The one-loop $B$-model or extended Hitchin model
appears to possess an anomaly with respect to changes in the
Kahler metric.

\acknowledgments V.P. would like to thank A. Dymarsky for useful
discussions. We would like to thank A. Neitzke and C. Vafa for
elucidating the conjecture of \cite{Dijkgraaf:2004te}.  We
understand that the gravitational anomaly has also been found some
years ago in unpublished work by A. Klemm and C. Vafa.  The work
of V.P. was supported in part by grant RFBR 04-02-16880 and grant
NSF PHY-0243680, and that of E.W. by NSF grant PHY-0070928.

\providecommand{\href}[2]{#2}\begingroup\raggedright\endgroup

\end{document}